\documentclass{aa}
\usepackage{graphicx}
\usepackage{natbib}
\bibpunct{(}{)}{;}{a}{}{,}

\begin{document}

\title{Confidence limits of evolutionary synthesis models}
\subtitle{III. On time-integrated quantities}

\author{M.~Cervi\~no\inst{1,2,3}
        \and M.A.~G\'omez-Flechoso\inst{4}
        \and F.J. Castander\inst{1,5,6}\thanks{Andes Prize Fellow.}
        \and D. Schaerer\inst{1}
        \and M. Moll\'a\inst{7}
        \and J.~Kn\"odlseder\inst{2} 
        \and V. Luridiana\inst{8,9}
}
\institute{UMR CNRS 5572, Observatoire Midi-Pyr\'en\'ees, 
              14, avenue Edouard Belin, 31400 Toulouse, France 
           \and Centre d'Etudes Spatiales des Rayonnements, 
              CNRS/UPS, B.P.~4346, 31028 Toulouse Cedex 4, France 
           \and Max-Planck-Institut f\"ur extraterrestrische Physik, 
              Giessenbachstrasse, 85748 Garching, Germany
           \and Observatoire de Gen\`eve, 
              CH-1290 Sauverny, Switzerland 
           \and Yale University, 
              P.O. Box 208101, New Haven, CT 06520-8101, USA 
           \and Universidad de Chile, 
              Casilla 36-D, Santiago, Chile 
           \and Departamento de F\'\i sica Te\'orica C-XI,
              Universidad Aut\'onoma de Madrid, 28049 Madrid, Spain 
           \and Instituto de Astronom\'\i a, UNAM, 
              Apdo. Postal 70-264, 04510 M\'exico D.F., Mexico
           \and European Southern Observatory, Karl-Schwarzschild-Str. 2, 
              D-85748 Garching bei M\"unchen, Germany
}

\offprints{Miguel Cervi\~no}
\mail{mcs@laeff.esa.es}
\date{Received: June 7, 2001 ; accepted: July 18, 2001 }
\authorrunning{M.~Cervi\~no et al.}
\titlerunning{Confidence Limits of Synthesis Models III}

\abstract{Evolutionary synthesis models are a fundamental tool to interpret
the properties of observed stellar systems. In order to achieve a
meaningful comparison between models and real data, it is necessary to
calibrate the models themselves, i.e. to evaluate the dispersion due to the
discreteness of star formation as well as the possible model errors. In
this paper we show that linear interpolations in the $\log M - \log t_k$
plane, that are customary in the evaluation of isochrones in evolutionary
synthesis codes, produce unphysical results.  We also show that some of the
methods used in the calculation of time-integrated quantities (kinetic
energy, and total ejected masses of different elements) may produce
unrealistic results.  We propose alternative solutions to solve both
problems.  Moreover, we have quantified the expected dispersion of these
quantities due to stochastic effects in stellar populations. As a
particular result, we show that the dispersion in the
$^{14}\mathrm{N}/^{12}\mathrm{C}$ ratio increases with time.
 } \maketitle

\section{Introduction}

Since their early introduction \citep{Tin80}, evolutionary synthesis models
have evolved to increase our understanding of the evolution of stellar
populations in the Universe.  The improvement of the observational
capabilities has forced the model developers to include more realistic
physical ingredients in the models (atmosphere models, grids of tracks
covering all the evolutionary phases, etc...)  to interpret the new data.
Thus, evolutionary synthesis models have become a useful tool to understand
the properties of observed stellar systems and to test the validity of
different evolutionary tracks. The improvements of synthesis models for
star-forming regions comprise mainly the inclusion of new physical inputs
and the extension of the output results to more observables.

Nevertheless, several aspects of synthesis models are still largely
perfectible.  An analysis of the way in which homology relations of massive
stars (implicitly used in evolutionary tracks interpolations and isochrone
computations) should be modified according to the assumed mass-loss rate,
and the inclusion of such new relations into the models, still need to be
performed. Additionally, the mathematical approximations used to estimate
the lifetimes of massive stars must be carried out carefully, otherwise
they could produce unphysical results.  Finally, the dispersion in the
model output parameters due to the discreteness of stellar populations has
been evaluated only in a few cases.

This work is the third paper of an on-going series whose the objective to
study the oversimplifications and the possible biases of the evolutionary
synthesis models for starburst regions, and to assess the confidence limits
of their outputs.  The global structure of the project is the following:
Paper {\sc i} \citep{CLC00} has been devoted to the study of the confidence
limits of synthesis models due to the discreteness of real stellar
populations. In Paper {\sc ii} \citep{CVGMH01} we have investigated the
Poissonian dispersion due to finite populations in non-time-integrated
observables, its quantitative evaluation and implementation on codes with
an analytical approximation of the Initial Mass Function (IMF), as well as
its relation with the dispersion in the output results of Monte Carlo
simulations.  In this paper we present an analytical {\it approximation} to
the Supernova rate (SNr) calculations in starburst galaxies and the
problems related with its determination in evolutionary synthesis codes. We
also study the influence of the interpolations in time-integrated
quantities (kinetic energy, $E_{kin}$, and total ejected masses of
different elements, $y_{z}$) and propose a more precise interpolation
technique in order to avoid the unphysical results obtained by the previous
models.  Some improvements of the interpolation techniques of evolutionary
tracks used in synthesis models will be presented in Paper {\sc iv}
(Cervi\~no 2001 in preparation). We will complete the series with a global
study of the expected dispersion as a function of different star-formation
laws (continuous and extended star formation).

In section \ref{sec:model} we present the evolutionary synthesis code
used. In Section \ref{sec:SN} we present an analytical estimate of the SNr
and how it is computed in evolutionary synthesis models. In Section
\ref{sec:Ek} we show how the released kinetic energy and the ejected masses
are computed, and we estimate their Poissonian dispersions and bias due to
different computation techniques. Finally we draw our conclusions in
Section \ref{sec:con}. All the results of this paper are available in
tabular form in our web server at {\tt http://www.laeff.esa.es/users/mcs/}.

\section{The evolutionary synthesis model}
\label{sec:model}

Since evolutionary synthesis calculations rely on the properties of stars
with far more mass values than available from stellar evolution
calculations, one possible source of error is the method used to
interpolate between the available stellar tracks.

A few works deal with this problem analytically, either proposing
analytical formulations for some phases of the stellar evolution
\citep{Touetal96}, or using an analytical population synthesis code
\citep{Pletal01}.  One advantage of analytical formulations is that the
functional dependence of the output quantities may be obtained.

In order to understand and quantify the errors introduced by synthesis
codes for star forming regions, we summarize the main characteristics of
synthesis models with non-analytical formulations and the methods used to
compute several parameters.

\begin{enumerate}

\item 
Most synthesis models interpolate in tables of evolutionary tracks.  Such
tables are discrete in their mass and time entries\footnote{In fact they
are discrete in the evolutionary sequence, i.e. each point in the table are
representative of a given evolutionary stage.}.  When one wants to obtain
the luminosity $L$, the effective temperature $T_\mathrm{eff}$, or other
properties, including quantities which change abruptly along the
evolutionary sequence (e.g.\ surface abundances), homology relations are
assumed to describe with sufficient accuracy the dependence of these
quantities with the initial stellar mass.  Therefore, interpolations in the
$\log M-\log A_k$ plane, where $M$ is the initial mass and $A_k$ is a
generic stellar property at a given evolutionary stage, are usually
performed.  Additional interpolations in the $\log M-\log t_k$ plane are
performed to obtain isochrones, and their validity will be examined in this
paper.  Whatever the approach is (fully analytical or table interpolation),
continuity of the stellar properties is assumed. The problem of
discontinuities in evolutionary tracks will be discussed in Paper {\sc iv}.

\item 
To calculate the integrated properties of the stellar population (e.g., the
luminosity in a given band) a numerical integration over the IMF-weighted
isochrones is always needed.  Two main approaches are used: either the IMF
is binned into a grid of $N$ initial masses, $M_i$, and then, all stars
belonging to the same mass bin are assumed to have {\it exactly} the same
properties, or the IMF is sampled with Monte Carlo simulations, and then
{\it each} individual star is evolved and the isochrone integration is
performed adding the evolved stars.

To avoid a bias due to the choice of the mass bin, a dynamical mass grid
can be used within the first method (\citealt{Mey95}, \citealt{SV98}, or
{\em Starburst99} \citealt{SB99}): at each computed age, the differences in
$L$ and $T_\mathrm{eff}$ between two stars of initial masses $M_i$ and
$M_{i+1}$, respectively, are constrained to be lower than a given
resolution $\Delta L$ and $\Delta T_\mathrm{eff}$.  Note that the resulting
mass grid will be different at different ages: the total number of bins $N$
and the $M_i$ values vary from one computed age to another (i.e., it is a
dynamical mass grid). Such method assures that the H--R diagram is mapped
in a continuous way and all the relevant evolutionary phases for the given
age (i.e. the isochrones) are included in the computations.

In the case of Monte Carlo simulations, either a high number of simulations
or a high number of stars in each individual simulation are needed to
produce a well mapped isochrone.  As a first order estimation, one
simulation with 10$^5$ stars in the mass range 2 -- 120 M$_\odot$ is
required \citep[as used in ][]{MHK91} to obtain Ultraviolet and optical
luminosities similar to those of analytical-IMF models. However, the
dispersion of Monte Carlo simulations depends on the considered observable.
Monte Carlo simulations have the advantage of allowing the straightforward
computation of the standard deviations and the confidence levels due to the
discreteness of the stellar population, provided the number of simulations
performed is high enough.  An additional advantage is that Monte Carlo
simulations take into account fast evolutionary phases that may be lost in
analytical simulations.  The required number of simulations needed to
obtain a satisfactory estimate of the dispersion can be estimated comparing
the mean values of the outputs with the results of analytical codes as far
as they must to coincide (at least for observables not related with fast
evolutionary phases).

In both cases, caution must be taken if the evolutionary tracks present
discontinuities (this subject is addressed in Paper {\sc iv}).

\item  
The final element of synthesis codes is the age resolution (or time step)
used to compute the integrated properties of simple stellar populations,
i.e.\ instantaneous bursts.  A sufficient temporal resolution, depending a
priori on the observable, is needed to assure accurate convolutions over
other arbitrary star-formation histories (e.g., the case of constant star
formation, subject addressed in Paper {\sc v}).

\end{enumerate}

For this study we have used the updated version of the evolutionary
synthesis code presented in \citet{MHK91,CMH94}. The updated code includes:

\begin{enumerate}
\item The full set of non-rotating Geneva evolutionary tracks including
standard \cite[ and references therein]{Schetal93} and enhanced mass-loss
rates \citep{Meyetal94}.
\item The metallicity-dependent atmosphere models for normal stars
from \citet{kur91}, the line blanketed non-LTE model atmospheres for O stars
({\it CoStar}, \citealt{SK97}) and the atmosphere models
for Wolf-Rayet (WR) stars from \citet{Schmetal92}.
\item A numerical isochrone integration using a modified dynamical mass bin,
now included in the Dec.\ 2000 release of Starburst99 \citep{SB99}. We
have also kept the original Monte Carlo formulation.
\item Parabolic interpolations in the $\log M - \log t_k$ plane for the
isochrones computation\footnote{During the work on the present models an
error, originating from the change from track to isochrone synthesis, was
found in the calculation of the SNr and some derived quantities in the {\em
Starburst99} code \citep{SB99}. This resulted in a strongly non-monotonic
SNr, partly increasing with time, in contrast with the expectations.  The
method described here has now been included in the December 2000 release of
the {\em Starburst99} code.} (see below).
\item The computation of the dispersion of all quantities for comparisons
with low-mass stellar systems (see Paper {\sc ii}).
\end{enumerate}

The calculations in the present paper are done with the solar metallicity
tracks of \citet{Schetal92}, adopting standard mass-loss rates, a Salpeter
IMF over the range from 2 to 120 M$_\odot$, and an instantaneous burst.

\section{The supernova rate}
\label{sec:SN}

The supernova rate, as any other output of an evolutionary synthesis code,
depends on the assumed IMF.  The study of the IMF has been broadly covered
in the astronomical literature \citep[see the volume of][]{Gil98}.  We
define the IMF as:

\begin{equation}
\Phi(M)=\frac{dN}{dM}=A\, M^{-\alpha}
\label{eq:imf}
\end{equation}

\noindent where $\alpha$ is the IMF slope, $A$ is a normalization factor
and $M$ the initial mass.  This function gives us the {\it probability} of
forming a number of stars in a given initial mass range.  The widely used
Salpeter's IMF slope corresponds to $\alpha = 2.35$ with this
definition. The number of stars $N_{star}(t)$ present in a system at a
given time from the onset of star formation within an initial mass range,
is obtained by the convolution of $\Phi(M)$ with the Star Formation Rate
law, $\Psi(t)$:

\begin{equation}
N_{star}=\int_{0}^{t} \int_{M_{\rm low}}^{M(t)} \Phi(M) \Psi(t-t') dM dt',
\label{eq:nstar0}
\end{equation}

\noindent where $M(t)$ is the initial mass of the star that ends its 
evolution at time $t$.

Throughout this work we assume an Instantaneous Burst (IB) of star
formation, $\Psi(t)=\delta(t)$ where $\delta(t)$ is Dirac's delta function,
so that:

\begin{equation}
N_{star}=\int_{M_{\rm low}}^{M(t)} \Phi(M) dM.
\label{eq:nstar}
\end{equation}

The number of stars that will end their evolution in the system\footnote{
This treatment is general for all stars described by the function
$M(t)$. Since we restrict ourselves to times shorter than 20 Myr, all the
stars considered will end their lives either with a SN explosion or with
the formation of a Black Hole.  Here we do not distinguish between these
cases.}, $N_{SN}$, in a time interval $[t_1,t_2]$ is

\begin{equation}
N_{SN}=\int_{M(t_{1})}^{M(t_{2})} \Phi(M) dM. 
\label{eq:nsn1}
\end{equation}

For mathematical convenience $M(t)$ can be approximated by

\begin{equation}
M(t) = B\, t^{-\gamma},
\label{eq:mt} 
\end{equation}

\noindent with $\gamma > 0$, thus assuming implicitly a linear relation
between $\log M$ and $\log t$.  Table \ref{tab:gamma} shows the values of
$\gamma$ and $\log B$ for several mass ranges from the \citet{Schetal92}
solar metallicity tracks using a linear $\log M - \log t$ approximation
(but see below). Eq. \ref{eq:nsn1} can be rewritten as:

\begin{equation}
N_{SN}=\int_{t_{1}}^{t_{2}} \Phi[M(t)] \left|\frac{dM}{dt}\right| dt
\label{eq:nsn2}
\end{equation}

Using Eq.~\ref{eq:nsn2} and \ref{eq:imf}, we obtain the SNr, i.e. the
number of SN in a time interval:

In the general case of a function $M(t)$, we obtain the {\it exact} value
of the SNr, i.e. the number of SN in a time interval:

\begin{equation}
SNr(t)= \frac{dN_{SN}}{dt} = A\, M(t)^{-\alpha}\,\left|\frac{dM}{dt}\right|. 
\label{eq:snr2}
\end{equation}

Using Eq.~\ref{eq:mt} (i.e. a linear interpolation in $\log M - \log t$)
one obtains:

\begin{equation}
SNr(t)= A\, B^{-\alpha+1}\,\,\gamma\,t^{\beta},
\label{eq:snr}
\end{equation}

\noindent where $\beta = \gamma\alpha-\gamma-1$.  For the Salpeter IMF
Eq.~\ref{eq:snr} shows that the SNr is a decreasing function of age.  This
expression is also useful to verify the proper calculation of the SNr in
evolutionary synthesis models that use linear interpolations in $\log M -
\log t$.

\begin{table}
\begin{tabular}{rcl|ccc}
\multicolumn{3}{c}{Mass range (M$_\odot$)} & $\gamma$ (Eq. \ref{eq:mt}) &
$\log B$ (Eq. \ref{eq:mt})& $\beta$ \\
\hline
 120  &--&  85   &  4.51 & 31.31 &  5.09\\
  85  &--&  60   &  1.88 & 14.15 &  1.54\\
  60  &--&  40   &  1.95 & 14.61 &  1.63\\
  40  &--&  25   &  1.21 &  9.68 &  0.63\\
  25  &--&  20   &  0.94 &  7.81 &  0.26\\
  20  &--&  15   &  0.81 &  6.96 &  0.10\\
  15  &--&  12   &  0.68 &  6.04 & -0.08\\
  12  &--&   9   &  0.57 &  5.22 & -0.23\\
   9  &--&   7   &  0.50 &  4.68 & -0.33\\
\hline
\end{tabular}
\caption{Values of $\gamma$ and $\log B$ defined in Eq. \ref{eq:mt} and the
SNr slope (assuming a Salpeter IMF slope) for different mass ranges from
the \citet{Schetal92} solar metallicity tracks.}
\label{tab:gamma}
\end{table}

\subsection{Implementation in evolutionary synthesis codes}

To calculate the SNr in evolutionary synthesis codes, we compute the
population at some given age, $t_j$.  The basic idea is to know how many
stars have ended their evolution between the previous computed age,
$t_{j-1}$, and the current one, $t_j$. Then, the SNr obtained is {\it the
mean value of the SNr for the used time interval}. At the age $t_j$ this is
given by:

\begin{equation}
SNr(t_j)=-\frac{\sum_{i=1}^N a_i
w_i}{t_{j}-t_{j-1}}=\frac{N_{SN}(t_j)}{t_j-t_{j-1}},
\label{eq:snrt} 
\end{equation}

\noindent where $w_i$ is the normalized number of stars of initial
mass\footnote{Note that for codes that use a dynamical mass binning, $w_i$
is in fact $w_i(t)$} $M_i$, and $a_i$ is defined as

\begin{equation}
a_i = \{ 
 \begin{array}{ll}
  0& \mathrm{if ~}i > i(t_j)  \mathrm{~ or ~} i < i(t_{j-1})\\
  1& \mathrm{if ~} i(t_{j-1}) \leq i \leq i(t_j)\\
 \end{array}
\label{eq:aiSN}
\end{equation}

\noindent where $i(t_{j})$ is the index in the binned IMF grid
corresponding to a mass $M(t_{j})$. The indexes are given by the function
$M(t)$ described above.

\begin{figure}
 \resizebox{\hsize}{!}{\includegraphics[angle=270]{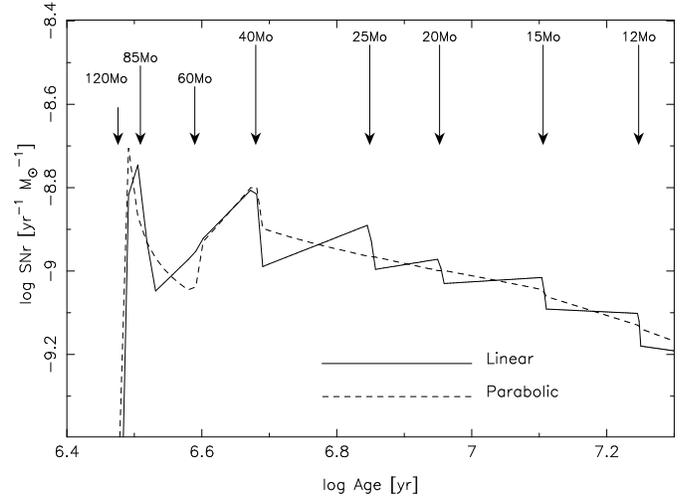}}
 \caption{SNr using different interpolation
 techniques. The solid line corresponds to a linear interpolation in the
 $\log M -\log t$ plane.  The short-dashed line corresponds to a
 parabolic
 interpolation.}
\label{fig:SN}
\end{figure}

The resulting SNr using a linear or parabolic interpolation of $M(t)$ is
shown in Fig.\ \ref{fig:SN}.  Whereas the SNr using the linear
interpolation (cf. Eq. \ref{eq:snr}) exhibits discontinuities corresponding
to the discreteness of the stellar tracks (cf.\ Table \ref{tab:gamma}), the
parabolic interpolation presents a much smoother behavior.  When linear
interpolations are used, the resulting lifetimes of stars at both sides of
a given tabulated stellar track are lager than the lifetimes obtained with
parabolic interpolations. This produces a lower SNr when the stars with
masses corresponding to the tabulated track had just exploded and an
accumulation of SN events just before the lifetime corresponding to the
following tabulated star.

\begin{figure}
 \resizebox{\hsize}{!}{\includegraphics[angle=270,width=7cm]{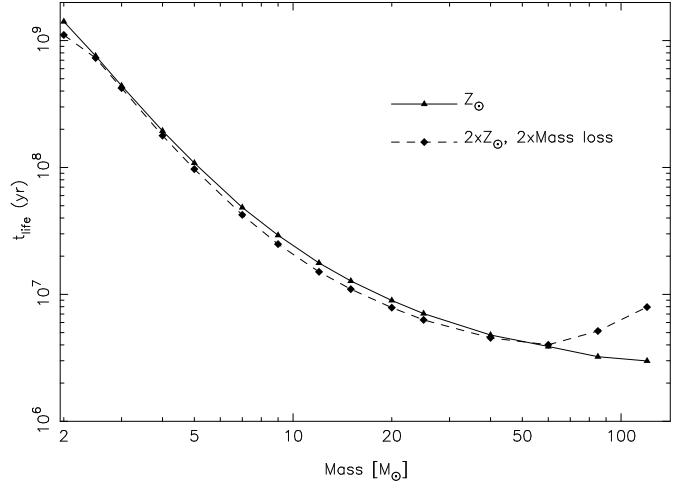}}
 \caption{$\log M - \log t_e$ plane for two different set of tracks:
 \citet{Schetal92} at solar metallicity and standard mass-loss rate and
 \citet{Meyetal94} at twice the solar metallicity and twice the mass-loss rate}
\label{fig:tlife}
\end{figure}

The situation is also illustrated in Fig \ref{fig:tlife} where the $\log M
- \log t$ plane is shown for the tracks used and the solar-metallicity and
twice solar-metallicity tracks of \citet{Meyetal94} (for masses lower than
12 M$_\odot$ we have completed the table with \citealt{Schetal93}).  A
general deviation from linearity is present for massive stars, such
deviation is more extreme in the case of twice solar metallicity tracks
with twice mass-loss rates.

It is interesting to remark that whatever the interpolation technique is,
some wiggles are seen at the beginning of the evolution of the SNr. Whereas
the abrupt discontinuities at the ages that correspond to the lifetime of
the 60 and 40 M$_\odot$ stars are due to the interpolation tecnique, the
wiggles themselves are due to the particular behavior of lifetime of the WR
stars in the set of tracks used. A detailed analysis of the lifetimes of
the tabulated stars that reach the WR phase reveals a nonmonotonic behavior
of the slope of the lifetime itself. This behaviour, convolved with the IMF
slope, produces the wiggles present in the figure; i.e., the presence of
the wiggles arise naturally given the set of tracks used, whereas the exact
ages where the wiggles appear are dependent on the interpolation technique.

The output of a synthesis code should not depend on the specific masses
tabulated in the evolutionary tracks (unless the tracks follow a
discontinuity of the stellar evolution). But the behavior of the
$\left|\frac{dM}{dt}\right|$ relation shows such dependence if a linear
interpolation in the $\log M - \log t$ plane is used. Moreover, the linear
behavior assumed in the $\log M - \log t$ plane is not real at all for some
set of tracks.  The misbehavior of linear interpolations may be due to the
effects of mass loss and overshooting in massive stars and it may also be
present in the new generation of tracks with rotation.

Finally, we remark again that we have focused on the ages of the SNe
explosions, but a similar situation exists at other evolutionary phases. We
want to stress that, even if the parabolic interpolation used here seem to
produce more realistic results, a correct interpolation technique (based on
physical principles) does not yet exist, and a more careful study is
necessary on this subject.  The parabolic interpolation subroutines
developed here are available at {\tt http://www.laeff.esa.es/users/mcs/}.

\subsection{Estimate of the SNr dispersion obtained from an 
evolutionary code}

The SNr is not a direct observable. However it enters in the calculation of
other observables like the non-thermal radio flux \citep{MHK91} or the
ejection of elements into the ISM. So, the knowledge of the SNr dispersion
due to the discreteness of the stellar population is needed to obtain the
expected dispersion in the observed properties of real systems.  In the
following paragraph we summarize how to calculate such a dispersion. We
refer to \citet{Buzz89} and Paper {\sc ii} for further details.

The IMF gives the probability, $w_i$, of finding a number of stars within a
given mass range at $t=0$. If we assume that each $w_i$ follows a
Poissonian distribution, the variance of each $w_i$, $\sigma_i^2$, is equal
to the mean value of the distribution, $w_i$.  Let us assume now that each
star has a property $a_i$, so that the contribution to the integrated
property $A$ of the star of the same mass is given by $w_i a_i$, with a
variance $\sigma_i^2 a_i^2 = w_i a_i^2$. The total variance of the
observable $A$ is the sum of all the variances. The relative dispersion is:

\begin{equation}
\frac{\sigma_{A}}{A}=
\frac{(\sum_{i=1}^N w_i a_i^2)^{1/2}}{\sum_{i=1}^N w_i a_i} =
\frac{1}{\sqrt{N_\mathrm{eff}(A)}}
\label{eq:neff}
\end{equation}

\noindent where the last term gives us the definition of
$N_\mathrm{eff}(A)$ described by \citet{Buzz89}.  Note that
$N_\mathrm{eff}(A)$ is normalized to the total mass, so $N_\mathrm{eff}(A)$
gives directly the relative dispersion for any total mass transformed into
stars.

Paper {\sc ii} shows that the dispersion obtained from Eq. \ref{eq:neff} is
equal to the dispersion of Monte Carlo simulations.  Let us stress that
such dispersion is present in Nature (star populations are always discrete
and finite) and it is not an evaluation of the errors of the synthesis
models, i.e. {\it the dispersion is also an observable}. This intrinsic
dispersion must be taken into account, before establishing any conclusion,
when fitting observed quantities to model outputs. Finally, the evaluation
of the dispersion depends on the interpolation techniques used, so a
correct interpolation technique is also required to fit this observed
property of Nature.

In the case of the $N_{SN}(t_j)$, $a_i$ is defined by Eq. \ref{eq:aiSN},
and the relative dispersion in $N_{SN}(t_j)$ is:

\begin{equation}
\frac{\sigma_{N_{SN}(t_j)}}{N_{SN}(t_j)}=\frac{1}{\sqrt{N_{SN}(t_j)}}
\label{eq:snsig}
\end{equation}

In this particular case ($a_i = 0$ or 1) the mean value and the variance
coincide as it is the case in Poissonian distributions.  The obtained
$N_{SN}$ is a mean value over the time step used, and the obtained
dispersion shows the variation about that mean value. The dispersion,
however, depends on how the mean value is computed, i.e. depends on the
time step used.

\begin{figure}
 \resizebox{\hsize}{!}{\includegraphics[angle=270,width=7cm]{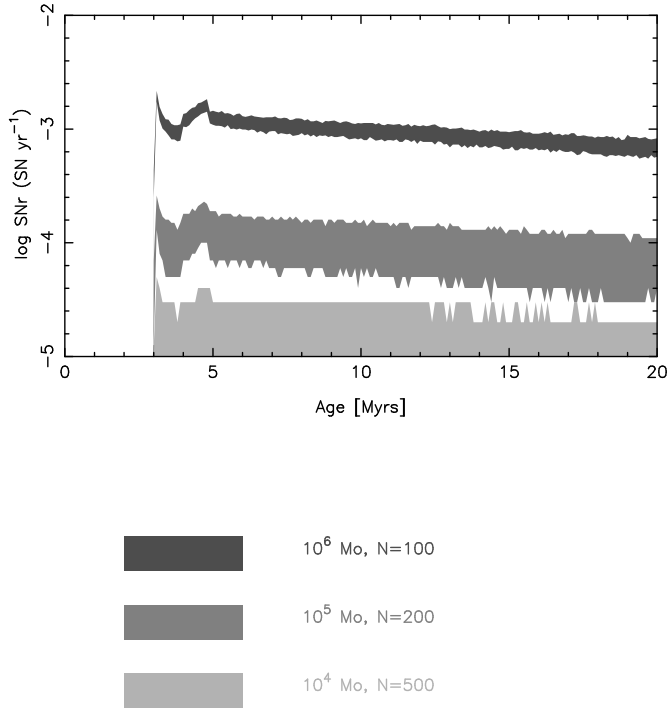}}
 \caption{90\% Confidence limit from Monte Carlo simulations for the SNr of
 clusters with different amount of total mass transformed into stars since
 the beginning of the burst of star formation with a time step of 0.1 Myr}
\label{fig:SNMon}
\end{figure}

The dispersion on the {\it mean} SNr is obtained dividing the variance
$\sigma^2_{N_{SN}}(t_j)$ by the time-step, i.e.
$N_\mathrm{eff}(SNr)=SNr(t_j)$.  Figure \ref{fig:SNMon} shows the 90\%
confidence limit for different Monte Carlo simulations. We have used 500
simulations of clusters with a total mass transformed into stars of 10$^4$
M$_\odot$, 200 simulations of clusters with 10$^5$ M$_\odot$, and 100
simulations of clusters with 10$^6$ M$_\odot$. A time step $\Delta t= 0.1$
Myr has been used in these simulations.

The dispersion obtained from the Monte Carlo simulations is given by
$\sigma_{Mon}$ = SNr(t) x $\Delta t$. This time-dependence in the
dispersion must be taken into account for quantities related to the
SNr. Note that the obtained dispersion is correct if the SNr is defined in
units of number of SN each $10^5$ years instead the usual units of SN per
year.

As an example, it is assumed that our Galaxy have a mean SNr about 3 SNe
per century which means, assuming a Poissonian distribution for the SNr,
$\sigma_{SNr} \sim 2$ SN per century and a relative dispersion of 0.6. If
the SNr is defined as 3 10$^4$ SN per Myr, the corresponding $\sigma_{SNr}$
becomes 173 SN per Myr and the relative dispersion is 0.006.

\section{Kinetic energy and ejected masses}
\label{sec:Ek}

The kinetic energy and ejected masses have two components: stellar winds
and SNe.  As we have pointed out before, all the outputs are affected by
the way the interpolations are performed in the $\log M - \log t_k$
plane. In the following we use parabolic interpolations in this plane.
Note that such interpolations produce a bit lower life-time that linear
ones, which means a lower amount of kinetic energy and integrated ejected
masses.  We now study these properties in more detail.

We assume a typical value of 10$^{51}$ erg SNe$^{-1}$ for the kinetic
energy released by a SN. The kinetic power, $P_{kin}(t)$, is the product of
the typical energy released by a SN multiplied by the $SNr(t)$.  In the
case of ejected masses, we need to use a relation between the SN ejected
masses and the mass of the exploding star. Illustrative examples can be
found in \citet{Poretal98} and \citet{Ceretal00}.

\subsection{Stellar wind components}

The kinetic energy and ejected masses also include a contribution from
stellar winds. The stellar mass-loss rate is the key parameter needed to
compute the kinetic energy and the ejected masses from stars before the end
of their evolution. The total kinetic power is the sum over the
contribution of individual stars, $p_i(t)$ at age $t$:

\begin{equation}
P_{kin}(t)=\sum_i^N w_i p_{i}(t) = \frac{1}{2} \sum_i^N w_i \dot{m_{i}}(t)
v^2_{\infty,i}(t),
\end{equation}

\noindent where $\dot{m_i}(t)$ is obtained from interpolations of the
tracks and $v_{\infty,i}(t)$ is derived from the interpolated luminosity,
the effective temperature, the mass $m_i$, and the metallicity of the star
of initial mass $M_i$ at the given age, following \citet{LH95}.

Similarly, the instantaneous ejected mass of an element $z$,
$\dot{y_z}(t)$, can be computed as:

\begin{equation}
\dot{y}_z(t)=\sum_i^N w_i \dot{y}_{z,i}(t) =\sum_i^N w_i \dot{m_i}(t)
Z_i(t),
\end{equation}

\noindent where $Z_i$ is the mass fraction of the surface abundance of the
element $z$ for a star of initial mass $M_i$ at age $t$. In general, in
evolutionary synthesis codes, interpolations in $L$, $T_{eff}$, $\dot{m}$
and $m$ are performed in the $\log M - \log A_k$ plane, and those in $Z$ in
the $\log M - Z_k$ plane, where the subindex $k$ refers to a given
evolutionary stage. In the case of chemical evolution models the
interpolations in $Z$ vary for different authors, from linear
interpolations in the $M - Z$ plane \citep{ferrini94} to the use of splines
\citep{leticia}.

For the computation of the corresponding {\it time integrated} quantities
--- the cumulative kinetic energy $E_{kin}$ and the total yield $y_z$ ---
an additional sum over time is needed.  Two different approximations may be
followed:

\begin{itemize}

\item {\it Method (a) The results from previous computed ages are used to
compute the sum:} Each value of $P_{kin}(t_j)$ or $\dot{y}_z(t_j)$ can be
considered constant over the time interval between $t_{j-1}$ and $t_j$,
where the index $j$ defines the age array used for the code output, and
the $E_{kin}$ and $y_z$ are obtained adding up such contributions. Note
that this method depends on the age array used and will lose evolutionary
stages where the characteristic time is shorter than the time step used.

\item {\it Method (b):} A similar treatment to those of chemical evolution
models is used. An additional table for each evolutionary state, $k$, with
the integrated amount of kinetic energy or chemical abundances from $t=0$
to each tabulated age, $t_k$ is computed. The final point of the table is
the integrated energy and ejected mass resulting from the action of stellar
winds all along the evolution of the star.

This method has two possible implementations:

\begin{itemize}
\item {\it Method (b.1):} The kinetic energy or chemical abundances tables
are obtained using the mass-loss, i.e. for the chemical abundances case:

\begin{equation}
y_{z,i}(t_{k'})=\sum_{k=1}^{k'} \dot{m_{i}}(t_k)\, Z_{i}(t_k)
\,(t_{i,k}-t_{i,k-1}),
\end{equation}

\noindent where the subindex $k$ refers to the tabulated points in the
tracks for the star of initial mass $M_i$. 

\item {\it Method (b.2):} The instantaneous mass is used instead of the
mass loss:

\begin{equation}
y_{z,i}(t_{k'})=\sum_{k=1}^{k'} [m_{i}(t_k)-m_{i}(t_{k-1})]\, <Z_{i}(t_k)>,
\end{equation}

\noindent where $<Z_{i}(t_k)>$ is the mean value of the surface abundance
at the evolutionary ages $t_{k}$ and $t_{k-1}$ of a star of initial mass
$M_i$. This is the approximation we have used here.

\end{itemize}

In both cases, the total amount of the ejected element $z$ at a given age
$t$ becomes:

\begin{equation}
y_z(t)=\sum_{i=1}^N w_i y_{z,i}(t)
\end{equation}

This method has the advantage that the output does not depend on the
age array used in the synthesis code.

\end{itemize}

The three methods should converge to the same value, but this can be only
achieved if the time step used in method (a) is the same as the lowest time
step used in the evolutionary tracks (i.e. a few years, that it is
prohibitive for realistic computations). Methods (b.1) and (b.2) must also
converge if

\begin{equation}
\dot{m_{i}}(t_k) \,(t_{i,k}-t_{i,k-1}) = m_{i}(t_k)-m_{i}(t_{k-1}),
\label{eq:mdot}
\end{equation}

\noindent which is, however, not found to be true for various sets of
stellar tracks adopted.

\begin{figure}
 \resizebox{\hsize}{!}{\includegraphics[angle=270]{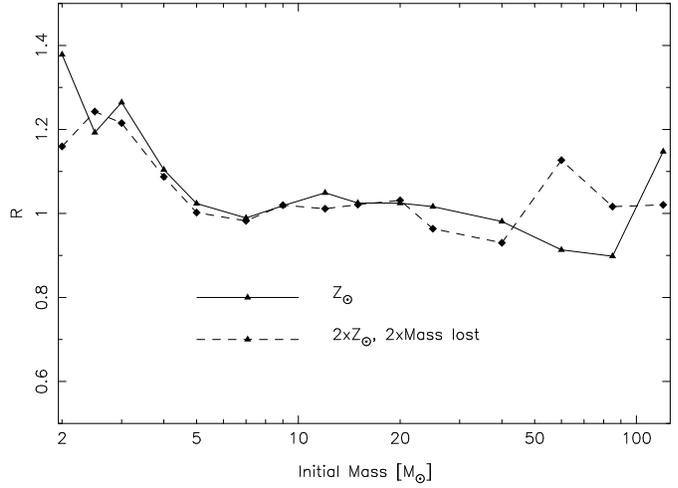}}
 \caption{Ratio of the integrated mass-loss during the lifetime of the star
 vs. ``reconstructed'' mass loss by subtraction of the mass at the Zero
 Age Main Sequence (ZAMS) and the mass at the end of the
 evolution for \citet{Schetal92} tracks (solid line) at solar metallicity
 and standard mass-loss rate and \citet{Meyetal94} tracks at twice solar
 metallicity and twice mass-loss rate tracks (dashed line).}
\label{fig:mdot}
\end{figure}

Let us illustrate the situation
defining the ratio $R$ for each star as:

\begin{equation}
R = \frac{\sum_{k=1}^{k_e} \dot{m}(t_k) \,(t_{k}-t_{k-1})}{M - m_{k_e}}
\end{equation}

\noindent where the index $k_e$ refers to the last tabulated point in the
track. Such ratio must be equal to 1 if Eq. \ref{eq:mdot} is fulfilled. The
resulting values are shown in Fig. \ref{fig:mdot} and in Table
\ref{tab:mdot} for the \citet{Schetal92} tracks at solar metallicity and
standard mass-loss rate and \citet{Meyetal94} tracks at twice solar
metallicity and twice mass-loss rate.

\begin{table}
\begin{tabular}{c|cc|cc}
Initial & $\sum \dot{m} \Delta t$ & $M-m_{k_e}$ & 
 $\sum \dot{m} \Delta t$ & $M-m_{k_e}$ \\
Mass range & (b.1) & (b.2) & (b.1) & (b.2) \\
(M$_\odot$) & \multicolumn{2}{c|}{Z$_\odot$} & 
\multicolumn{2}{c}{2xZ$_\odot$, 2x$\dot{M}$} \\
\hline
  12  & 0.5      &  0.4      & 2.3   & 2.3     \\
  15  &  1.5     &   1.4     & 3.5   & 3.5     \\
  20  &  3.6     &   3.5     & 10.1  & 9.8     \\
  25  &  9.6     &   9.4     & 18.6  & 19.3    \\
  40  &  31.3    &   31.9    & 33.0  & 35.5    \\
  60  &  47.6    &   52.2    & 63.3  & 56.2    \\
  85  &  68.3    &   76.0    & 83.8  & 82.5    \\
 120  &  129     &   112     & 120   & 118     \\
\hline
\end{tabular}
\caption{Values of the mass loss between the ZAMS and the end of the
 evolution,
and the integrated mass loss during the lifetime for different mass ranges
from the \citet{Schetal92} solar metallicity tracks and \citet{Meyetal94}
tracks at twice solar metallicity and twice mass-loss rate.}
\label{tab:mdot}
\end{table}

Typically, as shown in Fig.\ \ref{fig:mdot} and Table \ref{tab:mdot} the
ratio $R$ between the ``reconstituted'' integrated mass loss using the
above methods and the total mass loss given by the difference between the
tabulated initial and final mass of the tracks, is found to vary by up to
$\sim$ 10 \% for stars with non negligible mass loss.  Note that for a 120
M$_\odot$ star the integrated mass loss is equal or higher than the initial
mass!

\subsection{Global evolution and dispersion}

We now study the resulting integrated kinetic energy and chemical yields
including both stellar winds and SNe.  As an example, we have focused on
the ejected mass of $^{12}$C and $^{14}$N/$^{12}$C ratio.  These elements,
principally $^{14}$N, in a "standard" stellar population are mostly
produced by the intermediate stars in the Asymptotic Giant Branch (AGB)
phase and when they form Planetary Nebulae (PNe).  However AGB stars and
PNe appear at later ages than those discussed in our work.  Therefore, the
$^{14}$N and $^{12}$C produced in the first few Myr come only from massive
stars through stellar winds and SN explosions.  We have selected these two
elements as illustrative examples, to highlight the importance of the
WR-wind phase in massive stars.

\begin{figure}
 \resizebox{\hsize}{!}{\includegraphics[angle=270]{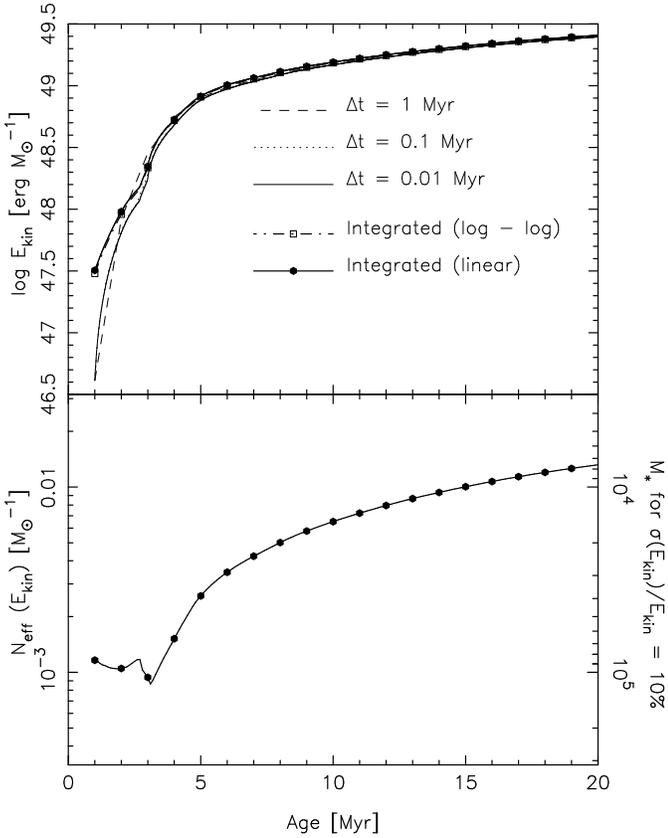}}
 \caption{{\it Top:} Integrated kinetic energy using three different time
 steps with method (a) and using method (b.2) described in the text with
 interpolations in the $M - E_{kin}$ plane (linear) and $\log M - \log
 E_{kin}$ plane (log -- log).  {\it Bottom:} $N_\mathrm{eff}(E_{kin})$ as
 defined in Eq. \ref{eq:neff} and computed by method (b.2) using the $M -
 E_{kin}$ plane for interpolations. The right vertical axis shows the
 minimum amount of gas that needs to be transformed into stars (in the
 given mass range and for the given IMF slope) to give a relative
 dispersion lower than 10\% when analytical-IMF models are compared with
 real data.}
\label{fig:tstep}
\end{figure}
\begin{figure}
 \resizebox{\hsize}{!}{\includegraphics[angle=270]{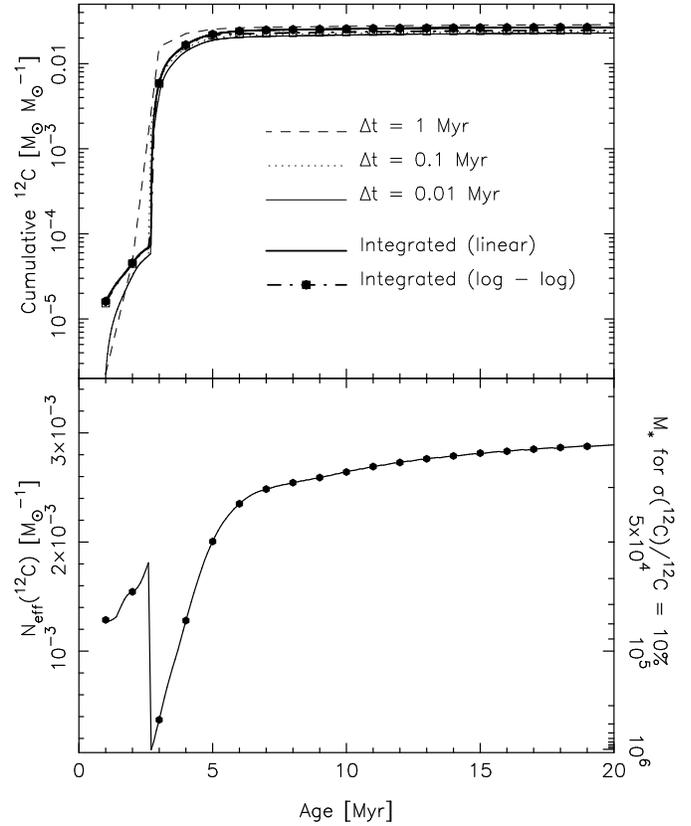}}
 \caption{Integrated amount of $^{12}$C ejected using three different time
 steps with method (a) and using method (b.2) described in the text with
 interpolations in the $M - ^{12}$C plane (linear) and $\log M - \log
 ^{12}$C plane (log -- log).  {\it Bottom:}
 $N_\mathrm{eff}(^{12}\mathrm{C})$ from method (b.2) using the $M - C $
 plane for interpolations. The right vertical axis shows the minimum amount
 of gas that needs to be transformed into stars (in the given mass range
 and for the given IMF slope) to give a relative dispersion lower than 10\%
 when analytical-IMF models are compared with real data.}
\label{fig:tstep2}
\end{figure}

\begin{figure*}
 \resizebox{\hsize}{!}{\includegraphics[angle=270,width=17cm]{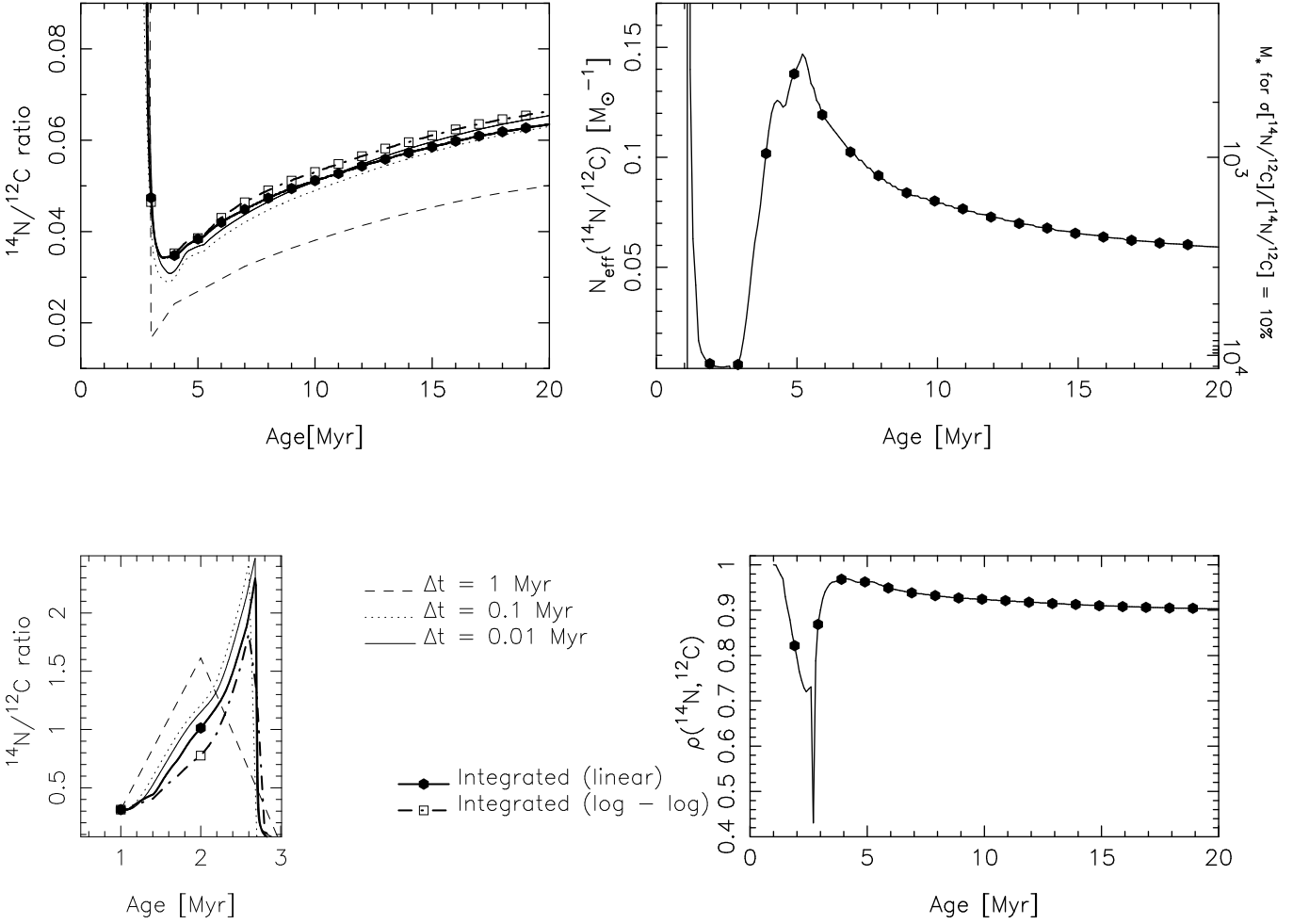}}
 \caption{{\it Top left:} $^{14}$N/$^{12}$C ratio obtained from the ejected
 $^{14}$N and $^{12}$C computed for three different time steps with method
 (a) and using method (b.2) described in the text with interpolations
 corresponding to Fig. \ref{fig:tstep2}.  {\it Bottom left:} Detail of
 $^{14}$N/$^{12}$C ratio obtained at the beginning of the burst of star
 formation. {\it Top right:} $N_\mathrm{eff}(^{14}\mathrm{N}/^{12}\mathrm{C})$
 (see text). The right vertical axis shows the minimum amount of gas that
 needs to be transformed into stars (in the given mass range and for the
 given IMF slope) to give a relative dispersion lower than 10\% when
 analytical-IMF models are compared with real data. {\it Bottom right:}
 Correlation coefficient, $\rho$($^{14}$N/$^{12}$C).} \label{fig:tstep3}
\end{figure*}

Figures \ref{fig:tstep}, \ref{fig:tstep2} and \ref{fig:tstep3} show these
quantities computed with method (a) for various time steps $\Delta t$, and
using method (b.2) with different interpolation techniques.  For method (a)
we use linear interpolations in $\log M - A$ for the abundances, and $\log
M - \log A$ for $E_{kin}$.  For method (b.2) we compare linear
interpolations in the $M - A$ and $\log M - \log A$ planes.  The
examination of the figures shows the following: {\em i)} time steps $\la$
0.1 Myr using method (a) appear adequate to properly calculate the
considered quantities; {\em ii)} at young ages dominated by stellar winds
($t \la$ 3 Myr), the use of pretabulated integrated quantities (method b)
leads to a somewhat larger and more correct values for kinetic energy and
yields produced by massive stars.  The origin of this difference is due to
rapid variations of these quantities along the isochrone, which are not
well enough resolved by method (a).  The differences between the different
numerical techniques are relevant for the younger ages, when the
integration of stellar winds are the only contribution of the
time-integrated quantities.  Using method (a) our code needs about 30' of
CPU time in a SunOS sparc machine using a time step of 0.1 Myr form 0.1 to
20 Myr. The CPU time is in this case inversely proportional to the time
step, i.e. the code would require about 300' of CPU time to obtain very
similar results once the SN activity is the dominant source, with a time
step of 0.01 Myr.  On the other hand, the time computations by method (b.2)
are not time-step dependent and they produce stable results whatever the
time step is.

In the lower panel of Fig.~\ref{fig:tstep}, \ref{fig:tstep2} and
\ref{fig:tstep3} we show the resulting $N_\mathrm{eff}$ for the kinetic
energy and ejected masses resulting from method (b.2) using interpolations
in the $M -A$ plane. $N_\mathrm{eff}$ can be easily computed for the
integrated properties if method (b.2) is used.  Note, though, that the
evaluation $N_\mathrm{eff}$ of the integrated properties from method (a)
with a dynamical mass binning is quite difficult because we need to know
the contribution of the {\it same individual} population along all the
computed ages.

In the case of the kinetic energy, the relative dispersion decreases with
age once the first SNe appears in the cluster. The natural explanation is
that more and more stars contribute to the released kinetic energy and the
statistics becomes better and better.

For the $^{12}$C case, there are three dominant contributions at different
times: massive non-WR stellar winds from 0 to 3 Myr, WR stellar winds from
3 to 5 Myr and the SNe thereafter. The depression in
$N_\mathrm{eff}(^{12}\mathrm{C})$ at 3 Myr coincides with the onset of the
WR phase. As it has been shown in Paper {\sc ii}, the WR phase is strongly
influenced by the discreteness of the stellar population and shows a high
dispersion in the related properties.  The cumulative production of
$^{12}$C by SNe decreases again the dispersion after 5 Myr.

The $^{14}$N/$^{12}$C ratio (Fig.\ \ref{fig:tstep3}, left panels) shows
several interesting effects.  It is characterized by an increasing ratio
until 3 Myr due to the effects of stellar winds of massive stars. WR stars
start to appear in the cluster at about 2 Myr. The evolution of WR stars at
solar metallicity follows the sequence of WR stars with N in the envelope
(WN phase, characterized by a strong mass-loss rate) and a posterior WC
phase, where C appears at the surface in an amount larger than N, with a
mass-loss rate dependent on the mass of the WR (i.e. decreasing with
time). The prevalence of massive OB stars and the WN phase last until 3 Myr
and produce more N than C. Later the action of WC winds and SNe increases
the C production.

The top right panel of Fig.\ \ref{fig:tstep3} shows the evolution of
$N_\mathrm{eff}(^{14}\mathrm{N}/^{12}\mathrm{C})$. The relative dispersion
(inverse of $N_\mathrm{eff}$) reaches a minimum value during the WR phase
as for $^{12}\mathrm{C}$. $N_\mathrm{eff}$ increases from the first SN
explosion to 5 Myr, and it decreases again slowly for more evolved
ages. The computation of $N_\mathrm{eff}(^{14}\mathrm{N}/^{12}\mathrm{C})$
must take into account the effect of the covariance (i.e. the correlation
coefficient, $\rho(^{14}\mathrm{N}/^{12}\mathrm{C})$).  Due to correlation
effects, the dispersion in the $^{14}$N/$^{12}$C ratio is lower than the
dispersion on $^{12}$C or $^{14}$N alone. Note also that the dispersion of
the $^{14}$N/$^{12}$C ratio increases as the cluster evolves.

\begin{figure}
 \resizebox{\hsize}{!}{\includegraphics[angle=270,width=7cm]{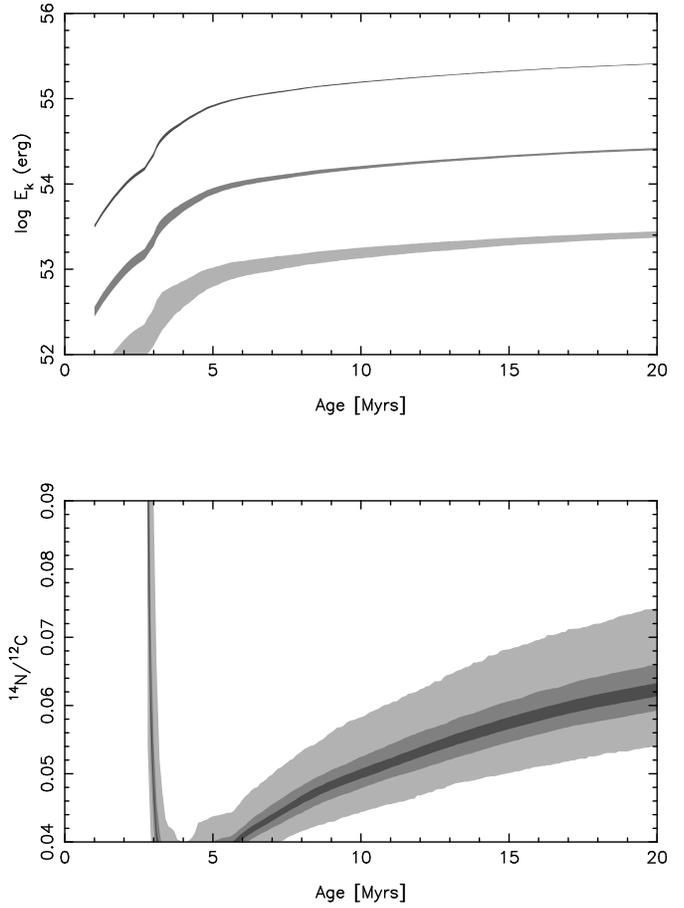}}
 \caption{90\% Confidence level of Monte Carlo simulations for $E_{kin}$
 and $^{14}\mathrm{N}/^{12}\mathrm{C}$ ratio of clusters with different
 amount of total mass transformed into stars since the beginning of the
 burst of star formation. The grey scale values are the ones defined in
 Fig. \ref{fig:SNMon}}
\label{fig:Mon}
\end{figure}

Finally Fig. \ref{fig:Mon} shows the resulting values of $E_{kin}$ and
$^{14}\mathrm{N}/^{12}\mathrm{C}$ ratio from a set of Monte Carlo
simulations. The figure shows the 90\% confidence level for simulations
where different amounts of gas have been transformed into stars. The exact
values of the 90\% confidence level can not be achieved by the numerical
formulation proposed, as far as the corresponding probability density
distributions are not known, but it is clear that the dispersion in the
simulations have a similar behavior than the one obtained by the
computation of $N_\mathrm{eff}$. In particular, Monte Carlo simulations and
the computation of $N_\mathrm{eff}(^{14}\mathrm{N}/^{12}\mathrm{C})$ show
that the dispersion in the $^{14}\mathrm{N}/^{12}\mathrm{C}$ ratio
increases with age.

\section{Conclusions}
\label{sec:con}

We have discussed technical issues (accuracy, impact of various
interpolation and integration methods) regarding the calculation of
supernova rates, as well as chemical and mechanical yields in evolutionary
synthesis models.  Furthermore we have quantified the expected dispersion
of these quantities due to stochastic effects in populations of various
total masses.  The main conclusions are the following:

\begin{enumerate}

\item

Linear interpolations in the $\log M - \log t_k$ plane, where $M$ is the
initial mass and $t$ is the age for a given evolutive stage, can give
unphysical results for the supernova rate in particular and for the 
evolutionary phases in general 

A parabolic interpolation technique can improve the results, but a more
general technique, taking into account the stellar evolution theory
(i.e. the relation of the stellar evolution parameters like luminosity and
effective temperature with time) would be an asset.  However, as far as
stellar theory is not complete, parabolic interpolations can be
used to produce more reasonable results than linear ones.

The unphysical results produced by the linear interpolations do not only
affect the massive stars (as naively expected), but also the low mass ones
(at least down to 9 M$_\odot$). The effects must be quantified for a wide
age range, including low mass stars, in the evolutionary synthesis models.

\item The time-integrated quantities of instantaneous burst models (or
single stellar populations) depend on the time step used for the
integrations. For the quantities considered here a time step not larger
than $\sim$ 0.1 Myr is found to be required for the integration of the
stellar winds component, that are relevant at the early phases of the
evolution of the cluster.  However, the best choice in terms of computing
time and accuracy is the use of predefined tables with the corresponding
time-integrated quantities.

\item In early phases ($t \la$ 3 Myr) the kinetic energy and the ejected
elements of star forming regions depend very strongly on the stellar
winds. In these phases, the most accurate outputs are obtained when
time-integrated tables of evolutionary tracks are used.

\item The discreteness of the real stellar populations is expected to
produce a dispersion in the observed parameters that must be taken into
account {\em a priori} when compared to the outputs of synthesis
models. For time-integrated quantities, the dispersion is higher at the
beginning of the star formation episode and becomes even more important
during the WR phase.  The dependence of the theoretical dispersion on the
total stellar mass has been quantified.

\item When the correlation between different yields of elements is taken
into account, the dispersion of the ratio of such elements increases with
time. The relevance of this effect on the observed dispersion of the
$^{14}\mathrm{N}/^{12}\mathrm{C}$ ratio remains to be evaluated.

\end{enumerate}

\begin{acknowledgements}
We want to acknowledge Claus Leitherer, Roberto Terlevich, Elena Terlevich,
Guillermo Tenorio-Tagle, Manuel Peimbert, Jes\'us Ma{\'\i}z Apellan{\'\i}z,
Jos\'e Miguel Mas-Hesse and David Valls-Gabaud for their useful comments
about this work. VL also acknowledges the Observatoire Midi-Pyr\'en\'ees
for providing facilities to conduct this study.  MC has been supported by
an ESA postdoctoral fellowship. MAGF is supported by the Direcci\'on
General de Ense\~nanza Superior (DGES, Spain) through a fellowship, FJC is
supported by an Andes Prize fellowship.
\end{acknowledgements}

\bibliographystyle{apj}
\end{document}